\documentclass[10pt,a4paper]{article}

\usepackage[utf8]{inputenc}
\usepackage[T1]{fontenc}
\usepackage{lmodern}

\usepackage[
    a4paper,
    left=1.7cm,
    right=1.7cm,
    top=1.6cm,
    bottom=1.6cm
]{geometry}

\usepackage{amsmath,amssymb,amsthm}

\usepackage{graphicx}
\usepackage{caption}
\usepackage{subcaption}
\usepackage{booktabs}
\usepackage{multirow}
\usepackage{array}
\usepackage{tabularx}

\usepackage{multicol}
\usepackage{titlesec}

\titlespacing*{\section}{0pt}{0.5ex}{0.1ex}
\titlespacing*{\subsection}{0pt}{1ex}{0.1ex}

\usepackage{natbib}

\usepackage{xcolor}
\usepackage[
    colorlinks=true,
    linkcolor=blue,
    citecolor=blue,
    urlcolor=blue,
    breaklinks,
    unicode=true, 
    psdextra      
]{hyperref}

\title{
\vspace{-0.5cm}
\textbf{Study of Butterfly-Shaped Coronal Hole Evolution Across the Solar Disk}
}

\author{
\textbf{Bhairab Ale$^{1,*}$, Khagendra Katuwal$^{2}$, and Shiv Narayan Yadav$^{1}$}
\\[0.15cm]
\small
$^{1}$\textit{Central Department of Physics, Tribhuvan University, Kirtipur, Nepal}
\\
$^{2}$\textit{Department of Astronomy, New Mexico State University, Las Cruces, USA}
\\
$^{*}$\textit{Corresponding author:}
\href{mailto:alebhairab0@gmail.com}{alebhairab0@gmail.com}
}
\date{}

\begin{document}

\maketitle


\begin{center}
\textbf{Abstract}
\end{center}
\textit{
We studied the evolution of a butterfly-shaped coronal hole observed during its passage across the solar disk from 2025 September 8 to 14. We used SDO/AIA 193~\AA\ images to identify the coronal-hole boundary and measure its area and mean intensity, and SDO/HMI line-of-sight magnetograms to examine the underlying photospheric magnetic field. The boundary was identified using a fixed intensity threshold of 100~DN, and the area, mean 193~\AA\ intensity, and total unsigned magnetic flux were tracked from the eastern to the western limb. As the coronal hole approached the central meridian, its measured area increased from approximately $7.70\times10^{16}$ to $1.63\times10^{17}~\mathrm{m}^{2}$ and its unsigned magnetic flux increased from $8.68\times10^{21}$ to $1.91\times10^{22}~\mathrm{Mx}$, while the mean intensity decreased from approximately $61$ to $45~\mathrm{DN}$. The area and unsigned magnetic flux followed similar temporal trends, whereas the mean 193~\AA\ intensity varied in the opposite direction. Near central-meridian passage, all three quantities remained relatively stable for several days, suggesting that the coronal-hole properties did not change abruptly during this interval. OMNIWeb observations detected a high-speed solar-wind stream approximately 2--4 days later, with speeds reaching $600$--$750~\mathrm{km~s^{-1}}$. The timing is consistent with the butterfly-shaped coronal hole being the likely solar source. These findings reveal a close relationship among coronal-hole area, 193~\AA\ intensity, and unsigned magnetic flux, and support an association with the subsequent high-speed solar-wind stream measured near Earth.
}
\vspace{0.3cm}

\noindent
\textbf{Keywords:} Coronal Hole, Unsigned Magnetic Flux, Solar wind, In-situ Observations, Velocity Fields, SDO/AIA $\&$ HMI


\begin{multicols}{2}

\section{Introduction}\label{sec:intro}
Coronal holes (CHs) are regions of the solar corona with lower density and temperature than the surrounding corona. They appear dark in extreme-ultraviolet (EUV) and X-ray observations \citep{Cranmer2009}. Their magnetic field is mainly open \citep{2023SPD....5420402K,2023AAS...24221404K}, allowing plasma to escape into interplanetary space. For this reason, CHs are important source regions of high-speed solar-wind streams \citep{ Zurbuchen2007,2024SoPh..299...54W,2026AAS...24743004K}.

CHs are generally classified as polar or equatorial. Polar CHs are large and long-lived structures commonly observed near the solar poles, especially during solar minimum. They can produce solar-wind speeds of about 700--800~km~s$^{-1}$ \citep{hofmeister2017characteristics,2024IAUS..365..383A}. Equatorial CHs occur at lower latitudes and can rotate across the Earth-facing side of the Sun. Their high-speed solar-wind streams can therefore influence the near-Earth space environment.

The evolution of a CH is closely related to the photospheric magnetic field. Although CHs are mainly unipolar, they can contain small-scale mixed-polarity magnetic features \citep{2023SPD....5420402K, 2026ApJ...999...63K,2025AAS...24611005K}. Magnetic-flux emergence, cancellation, transport, and reconnection can change the CH boundary, area, and magnetic structure. Studying these changes is important for understanding how CHs evolve and how they contribute to the solar wind.

Identifying the exact boundary of a CH remains difficult. CHs are commonly detected from their low EUV intensity, but the intensity boundary does not always match the region of open magnetic field \citep{Krista2009, Cranmer2009}. Measurements of CH area and magnetic flux can also depend on the selected intensity threshold, image-processing method, and projection effects \citep{Krista2009, hofmeister2017characteristics}. A consistent boundary-detection method is therefore needed when tracking a CH over time.

In situ observations provide direct measurements of the solar wind and interplanetary magnetic field near Earth. High-speed solar wind from CHs usually has higher speed and temperature and lower proton density than the slow solar wind \citep{Cranmer2009, zurbuchen2012sources,2026AAS...24743004K}. Comparing these measurements with solar observations helps connect the evolution of a CH with its effects in the heliosphere.

In this study, we investigate a long-lived, butterfly-shaped CH observed in September 2025. We track its area, AIA 193~\AA{} intensity, boundary structure, and photospheric magnetic properties as it crosses the solar disk. We then compare these changes with near-Earth solar-wind plasma, interplanetary magnetic field, and geomagnetic observations to study the connection between the CH and its heliospheric signatures.


\subsection{Data Set}
\label{sec:dataset}
We investigated a long-lived, butterfly-shaped coronal hole (CH) observed on the solar disk from 8 to 14 September 2025. This interval captures its evolution during the disk passage, including its transit across the central meridian. Observations were selected at eight times per day---03:00, 06:00, 09:00, 12:00, 15:00, 18:00, 21:00, and 23:59~UTC---providing an approximately three-hour cadence for monitoring changes in the CH morphology and magnetic properties. The analysis was designed to examine how the temporal evolution of the CH area, boundary, and magnetic configuration was related to the associated high-speed solar-wind stream and its signatures in near-Earth space.

\subsection{Atmospheric Imaging Assembly (AIA)} \label{sec:aia}

The Atmospheric Imaging Assembly (AIA; \citep{lemen2012atmospheric}) onboard the \textit{Solar Dynamics Observatory} provides near-simultaneous, full-disk observations of the solar atmosphere in multiple wavelength channels. In this study, we used images from the 193~\AA{} channel, which is primarily dominated by Fe~\textsc{xii} emission formed at a characteristic temperature of approximately $\log T \approx 6.1$.  The AIA 193~\AA{} images have a temporal cadence of 12~s and a pixel scale of approximately $0.6^{\prime\prime}$~pixel$^{-1}$. These observations were used to follow the evolution of the coronal hole throughout its disk passage and to quantify changes in its projected area, boundary morphology, and mean intensity. The derived coronal-hole properties were subsequently compared with the corresponding photospheric magnetic-field measurements.

\subsection{Helioseismic and Magnetic Imager (HMI)} \label{sec:hmi}
The Helioseismic and Magnetic Imager (HMI; \citep{schou2012design}) onboard the \textit{Solar Dynamics Observatory} provides continuous, full-disk observations of the solar photosphere using the Fe~\textsc{i} 6173~\AA{} absorption line. HMI measures the polarization and wavelength dependence of this spectral line to derive photospheric magnetic-field and Doppler-velocity data products. In this study, we used the 45~s line-of-sight magnetic-field data, $B_{\mathrm{LOS}}$, with a pixel scale of approximately $0.5^{\prime\prime}$~pixel$^{-1}$. The HMI observations were used to examine the photospheric magnetic-field distribution associated with the coronal hole and its temporal evolution during the selected observing period.

\subsection{In-situ Data} \label{sec:insitudata}
Near-Earth solar-wind and interplanetary magnetic-field measurements were obtained from the OMNIWeb database \footnotemark.
\footnotetext{\url{https://omniweb.gsfc.nasa.gov/}}. We analyzed the solar-wind bulk speed, $V$, and its components ($V_x$, $V_y$, and $V_z$), together with the proton number density, $N_p$, proton temperature, $T_p$, and dynamic pressure, $P_{\mathrm{dyn}}$. The interplanetary magnetic field (IMF) was characterized using its Cartesian components ($B_x$, $B_y$, and $B_z$) and total magnitude, $B$. The SYM-H index was also examined to assess the corresponding geomagnetic response. The OMNI measurements are propagated to the nominal nose of Earth’s bow shock, providing a time-consistent dataset for investigating the solar-wind and geomagnetic signatures associated with the coronal-hole high-speed stream.

\begin{table*}[ht!]
	\centering
	\caption{Key characteristics of the remote-sensing and in-situ datasets used in this study.}
	\label{tab:datasets}
	\small
	\begin{tabular}{p{2cm} p{2.25cm} p{3.25cm} p{3.25cm} p{3.25cm}}
		\hline
		\textbf{Instrument} & \textbf{Data Type} & \textbf{Observed Quantity} & \textbf{Wavelength/Channel} & \textbf{Cadence/Resolution} \\
		\hline\\
		HMI (SDO) & Remote-sensing & Photospheric magnetic field & Fe I 6173~\AA & 45 sec; $0.5''$ per pixel \\
		AIA (SDO) & Remote-sensing & Coronal EUV intensity & 193~\AA\ (Fe XII, $\log T \approx 6.1$) & 12 sec; $0.6''$ per pixel \\
		OMNIWeb & In-situ & Solar wind plasma \& IMF & -- & 5 min \\
		\hline
	\end{tabular}
\end{table*}

\section{Methodology}
\label{sec:methodology}
The AIA 193~\AA{} image shown in the left panel of Figure~\ref{fig:ch_11sep} and the HMI 45~s line-of-sight (LOS) magnetogram shown in the right panel were obtained from the Joint Science Operations Center (JSOC). For each observing time, we selected an AIA image and the HMI magnetogram closest in time. The AIA image was reprojected onto the HMI coordinate system using the \texttt{reproject} package, placing both data sets on the same spatial grid. We then cropped the coronal-hole region from both observations, as indicated by the black dotted box in Figure~\ref{fig:ch_11sep}. The coronal-hole boundary was identified from the AIA 193~\AA{} image using a fixed intensity threshold of 100~DN, following the method of Katuwal and McAteer \citep{2023SPD....5410302M,katuwal_2025_16521356,2026ApJ...999...63K,beck2025chasmswpcdatasetcoronalhole}. Pixels with intensities below this threshold were classified as part of the coronal hole. The same threshold was applied at all observing times to identify the boundary consistently. The detected boundary is shown on the AIA 193~\AA{} image in the left panel of Figure~\ref{fig:ch_11sep} and is overlaid on the HMI LOS magnetogram in the right panel.

\begin{figure*}[ht!]
	\centering
	\includegraphics[width=\textwidth]{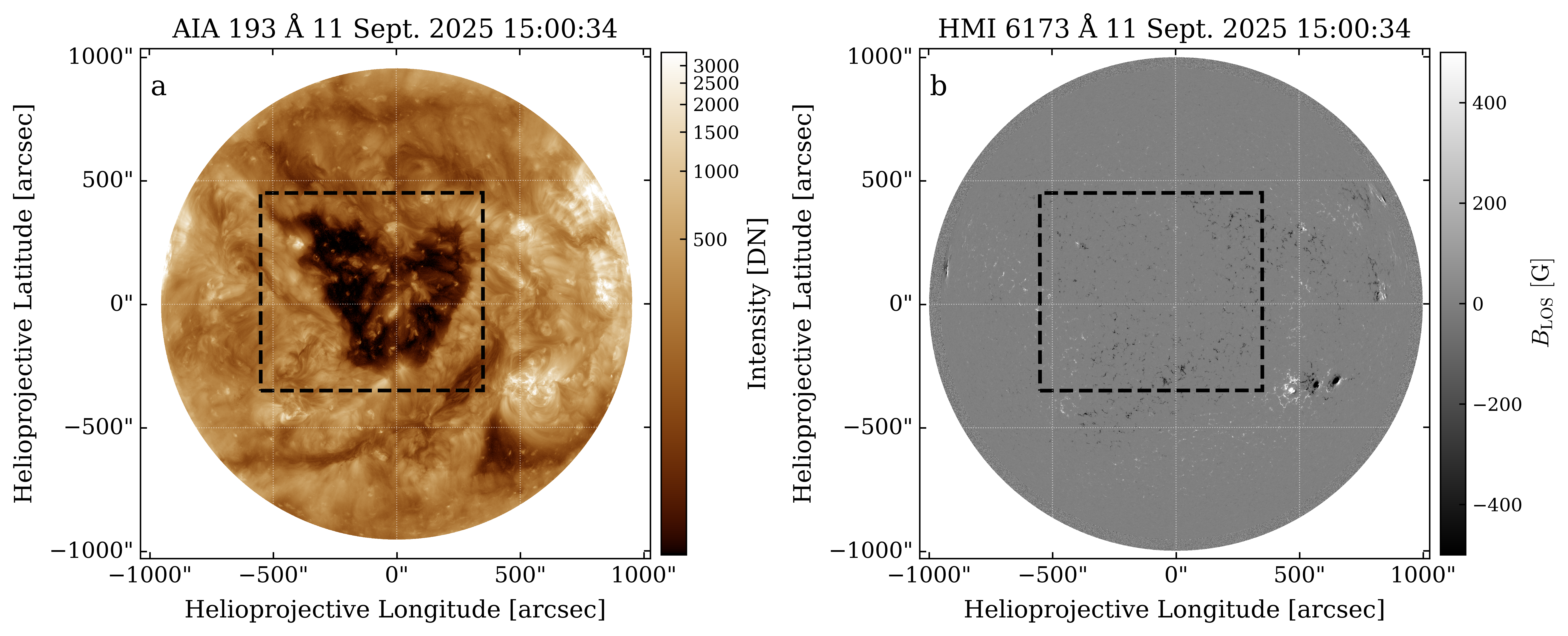}
	\caption{Reprojected AIA 193~\AA\ image and corresponding HMI magnetogram of the selected coronal hole on 11 September 2025 at 15:00:34. Panel (a) shows the CH at disk center, while panel (b) displays the associated line-of-sight magnetic field (\(B_{\mathrm{LOS}}\)). The black dotted box marks the region of interest.}
	\label{fig:centerday}
\end{figure*}

\begin{figure*}[ht!]
	\centering
	\includegraphics[width=\textwidth]{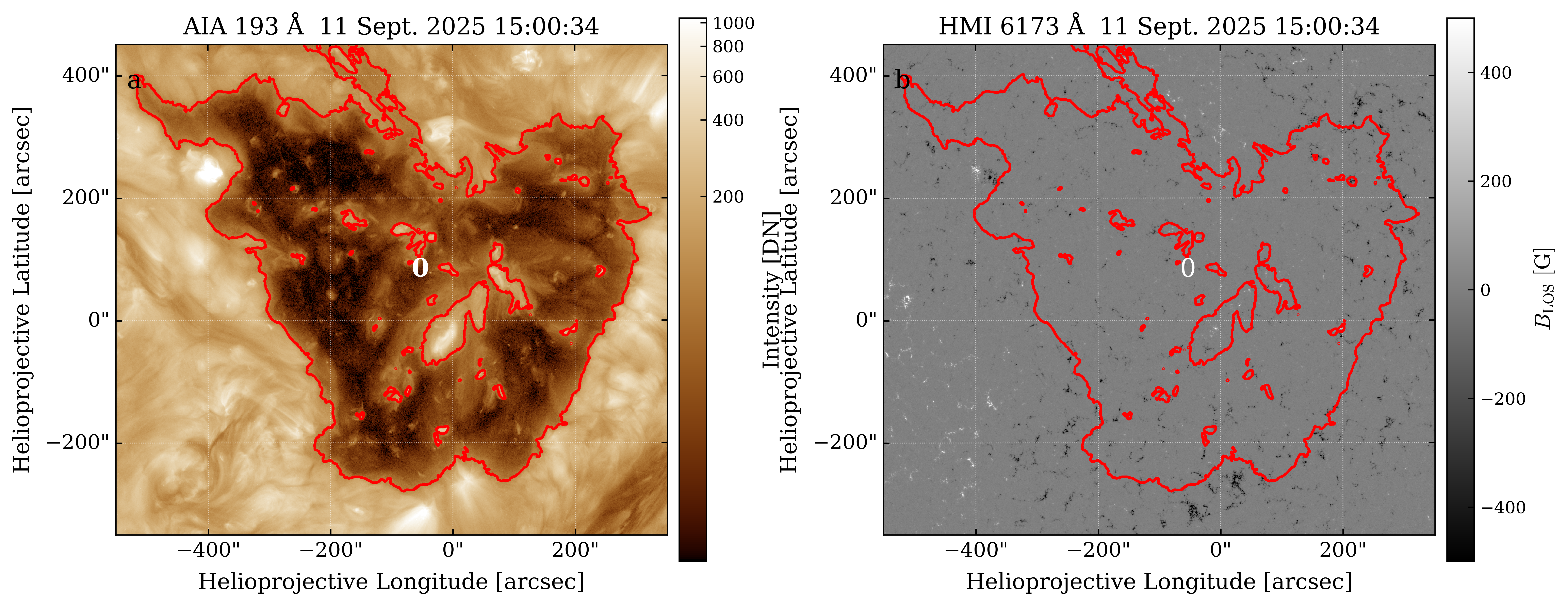}
	\caption{Coronal hole boundaries identified using a single intensity threshold of 100~DN on 11 September 2025 at 15:00:34 (Figure~\ref{fig:centerday}). Panel (a) shows the EUV image with the extracted boundary, and panel (b) shows the corresponding magnetogram in gray with the same boundary overlaid.}
	\label{fig:ch_11sep}
\end{figure*}

The coronal-hole area was calculated from the number of pixels inside the detected boundary. We corrected the pixel area for solar foreshortening using $\mu_i=\cos\theta_i$, where $\theta_i$ is the heliocentric angle of pixel $i$. The corrected area was calculated as
\begin{equation}
A_{\mathrm{CH}}
=
\sum_i M_i \frac{A_{{\rm pix},i}}{\mu_i},
\label{eq:ch_area}
\end{equation}
where $M_i$ is the coronal-hole mask, with $M_i=1$ inside the boundary and $M_i=0$ outside, and $A_{{\rm pix},i}$ is the projected area of pixel $i$. The area was expressed in units of ${\rm M}^{2}$.

The mean AIA 193~\AA{} intensity inside the coronal hole was calculated as
\begin{equation}
\langle I_{193}\rangle
=
\frac{\sum_i M_i I_{193,i}}
{\sum_i M_i},
\label{eq:mean_intensity}
\end{equation}
where $I_{193,i}$ is the intensity of pixel $i$.

The total unsigned LOS magnetic flux inside the same boundary was calculated from the HMI magnetograms as
\begin{equation}
\Phi_{\text{unsigned}}
=
\sum_i M_i
\left|B_{\mathrm{LOS},i}\right|
\frac{A_{{\rm pix},i}}{\mu_i},
\label{eq:unsigned_flux}
\end{equation}
where $B_{\mathrm{LOS},i}$ is the LOS magnetic field at pixel $i$. The area correction accounts for the changing viewing angle as the coronal hole moves across the solar disk \cite{hofmeister2017characteristics}. We applied the same procedure at each observing time to track changes in the coronal-hole area, mean 193~\AA{} intensity, boundary shape, and total unsigned magnetic flux.

Finally, we compared these remote-sensing measurements with near-Earth in situ data from the OMNIWeb database. We examined the solar-wind speed, proton density, proton temperature, dynamic pressure, interplanetary magnetic-field components, total magnetic-field strength, and the SYM-H index. The related solar-wind signatures reached Earth approximately 2--3~days after the coronal hole was observed on the Sun. This comparison allowed us to study the connection between the evolution of the coronal hole and the plasma, magnetic-field, and geomagnetic changes measured near Earth.

\section{Results and Discussion}
\label{sec:resultndis}
\subsection{Variation in $\langle I_{193 ~\AA\ }\rangle$ Across the Solar Disk}
\label{sec:aia_analysis} 
By measuring the average AIA 193~\AA\ intensity inside the extracted coronal-hole boundary, we identified variations in EUV emission as the coronal hole moved across the solar disk.

\begin{figure*}[ht!]
	\centering
	\includegraphics[width=0.8\textwidth]{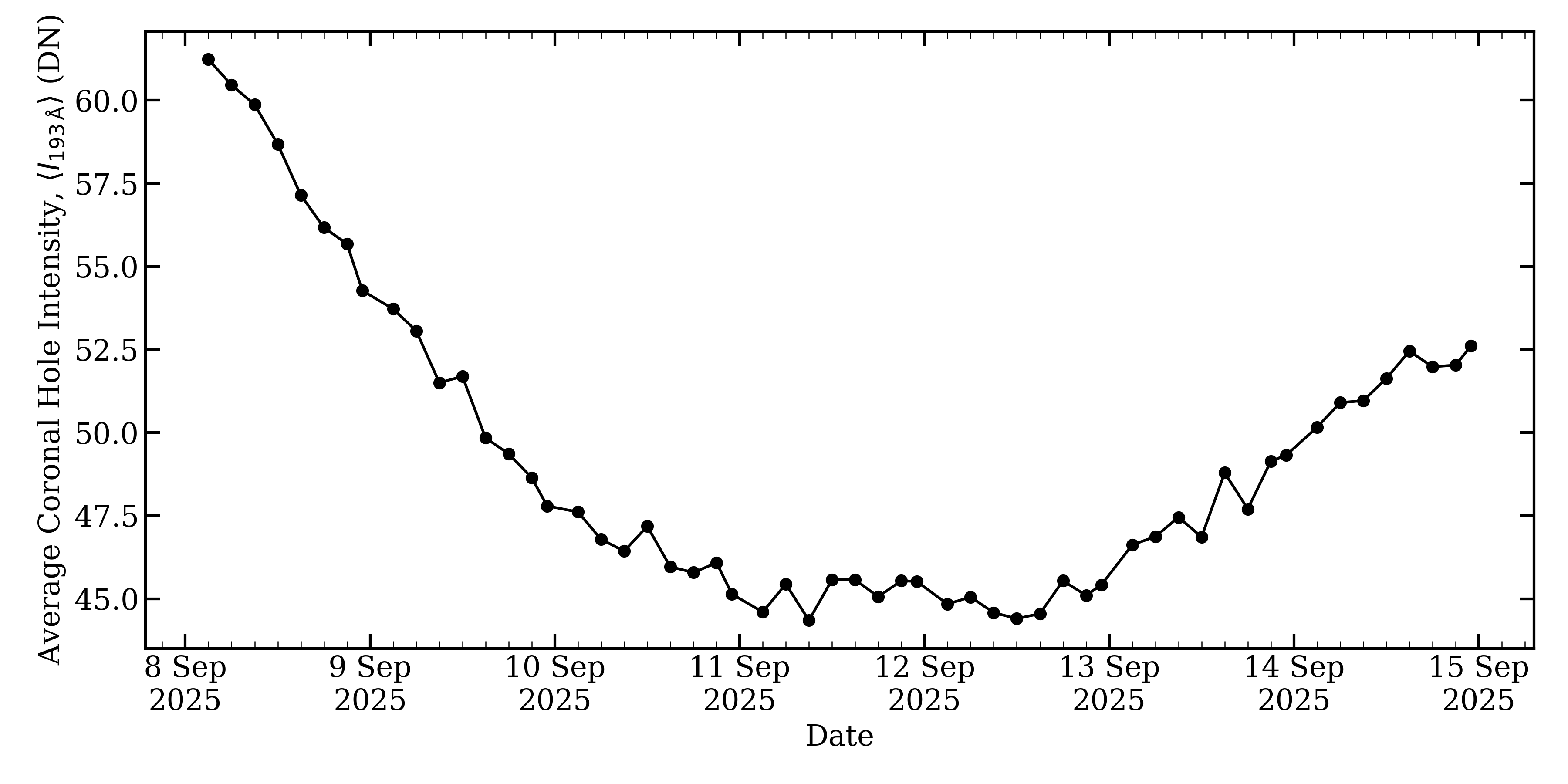}
	\caption{Average coronal hole EUV intensity ($\langle I_{193\,\mathrm{\AA}}\rangle$; DN) measured along the coronal hole boundaries. The plot shows the mean intensity within the 100 DN mask at a 3 hour cadence from 8 to 14 September 2025, reflecting variations as the feature transits the solar disk.}
	\label{fig:aia_result}
\end{figure*}

As shown in Figure~\ref{fig:aia_result}, the data demonstrate a strong evolutionary pattern in intensity from 8--14 September 2025, associated with the coronal hole rotation. On September 8, with the coronal hole located near the eastern limb, the average intensity peaked at 61.23 DN at 03:00 UT before gradually declining to 54.27 DN by 23:00 UT. This downward trend continued until 10 September, reaching 45.14 DN at 23:00 UT as the feature rotated toward the central meridian, obtaining a steady minimum of 44--45 DN from 11 to 12 September.

From 13 September onward, the plot indicates a gradual increase, with values increasing back to 50.16--52.61 DN by 14 September as the feature approached the western limb. Observations show that, while the mean intensity remains constant around the disk center, it changes dramatically closer to the edges of limbs. Variations of this size over such a short period are physically unreliable; hence, these changes are most likely the result of foreshortening effects driven by the partial visibility of coronal hole regions when projected near the limb.

We find that the average AIA 193~\AA\ intensity of the coronal hole remains nearly constant near the solar disk center. This nearly unchanged intensity indicates that the coronal hole maintains a stable structure over several days, consistent with the morphological stability reported by \citep{2023A&A...679A.100H,2026ApJ...999...63K}.

\subsection{Variation in $\Phi_{\text{unsigned}}$ Across the Solar Disk}

To study how the magnetic field changed with intensity, we estimated the total unsigned magnetic flux within the coronal hole parts. As indicated in Section~\ref{sec:methodology}, we detected the coronal hole borders using a 100 DN threshold on AIA 193~\AA\ pictures and overlaying these contours on corresponding SDO/HMI line-of-sight magnetograms. At each time step, we added the absolute value of the line-of-sight magnetic field ($B_{\rm LOS}$) for each pixel within the border. \cite{hofmeister2019photospheric} propose a polarity-independent estimate of total photospheric flux, capturing contributions from both positive and negative magnetic signatures.

\begin{figure*}[ht!]
	\centering
	\includegraphics[width=0.8\textwidth]{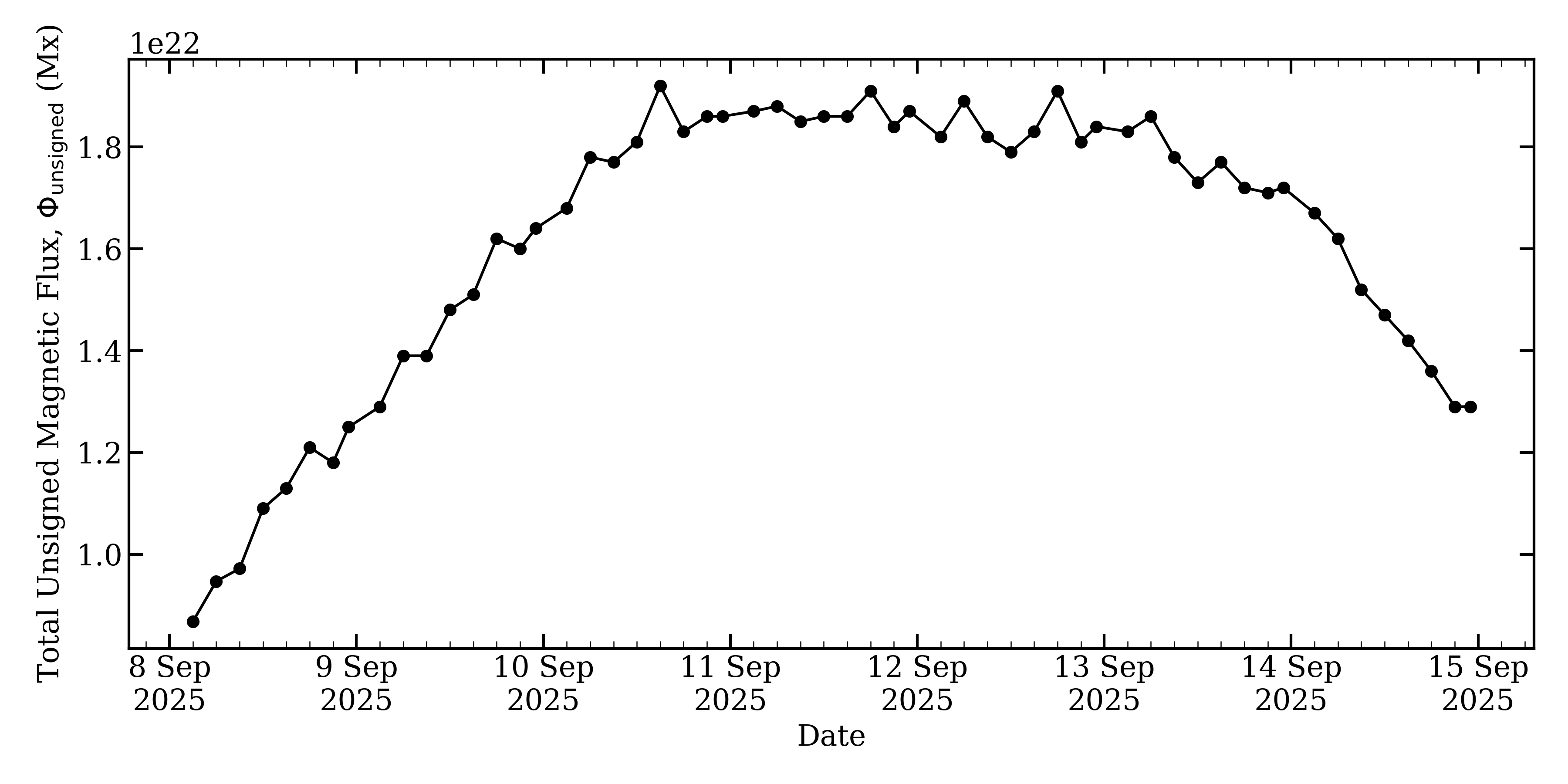}
	\caption{ Total unsigned magnetic flux ($\Phi$) integrated over the masked coronal hole area. The y-axis is scaled in units of $10^{17}$ Mx. Data points represent the 3-hour average of SDO/HMI magnetogram observations.}
	\label{fig:hmi_result}
\end{figure*}

The temporal data in Figure~\ref{fig:hmi_result} show a characteristic three-phase evolution of magnetic activity: an early spike, a stable period, and finally a fall. This magnetic behavior closely reflects the spatial contours given in Figure~\ref{fig:ch_11sep}, allowing us to analyze how the magnetic environment changes as the coronal hole transited the solar disk.

On September 8, the unsigned flux began at $8.68 \times 10^{21}$~Mx at 03:00~UT and gradually increased to $1.25 \times 10^{22}$~Mx by the end of the day at 23:00~UT. This early spike occurred as new magnetic features appeared and existing flux rotated into view from the sun's eastern side. This growing pattern remained until 10 September, with the flux reaching $1.86 \times 10^{22}$~Mx at 23:00~UT.

Between September 11 and 12, the coronal hole reached relative magnetic equilibrium. The unsigned flux remained stable, varying only slightly between $1.79 \times 10^{22}$~Mx and $1.91 \times 10^{22}$~Mx. However, the trend shifted to a slow fall beginning on September 13. The flux declined from $1.83 \times 10^{22}$~Mx at 03:00~UT to $1.72 \times 10^{22}$~Mx by the end of the day at 23:00~UT, eventually reducing to $1.29 \times 10^{22}$~Mx on September 14th at 23:00~UT.

From the above observation, we found that the total unsigned magnetic flux remained remarkably stable during the central meridian passage. According to Katuwal et al. \cite{2026ApJ...999...63K}, this stability in an equatorial coronal hole suggests a strong magnetic unipolarity. Even in the presence of surrounding mixed-polarity fields, the 'butterfly' structure maintains its integrity through a dominant open-field topology, which prevents significant flux cancellation at the boundaries.

\subsection{Variation in $A_{\mathrm{CH}}$ Across the Solar Disk}
The evolution of the coronal hole (CH) was analyzed using AIA 193 Å images from 8–14 September 2025. The CH area was determined by overlaying the boundaries identified with a 100 DN threshold on AIA 193 Å images, as described in Section~\ref{sec:methodology}. On 8 September, the measured area was $7.70 \times 10^{16}$ m$^2$ at 03:00~UT and increased steadily throughout the day, reaching $1.11 \times 10^{17}$ m$^2$ by 23:00~UT. This initial growth indicates that the CH was expanding and becoming more clearly defined as it rotated onto the visible solar disk.

The expansion continued on 9--10 September, with the area increasing from $1.15 \times 10^{17}$ m$^2$ to a maximum of $1.63 \times 10^{17}$ m$^2$ late on 10 September. This period represents the full development of the CH, when its spatial extent was largest and the boundaries were most clearly detected.

From 11--12 September, the area remained relatively stable, fluctuating slightly between $1.54 \times 10^{17}$ and $1.63 \times 10^{17}$ m$^2$, suggesting a temporary equilibrium. On 13 September, it decreased from approximately $1.57 \times 10^{17}$ m$^2$ at 03:00~UT to $1.45 \times 10^{17}$ m$^2$ by 23:00~UT, and by 14 September, it further reduced to $1.07 \times 10^{17}$ m$^2$ at 23:00~UT. This reduction reflects the gradual weakening and dispersal of the CH as it rotated toward the western limb, compounded by projection effects near the solar limb. Our analysis confirms a strong positive correlation between these parameters \textit{i.e.} Coronal Hole expansion is characterized by a decrease in EUV intensity and a concurrent increase in total unsigned magnetic flux.

\begin{figure*}[ht!]
	\centering
	\includegraphics[width=0.8\textwidth]{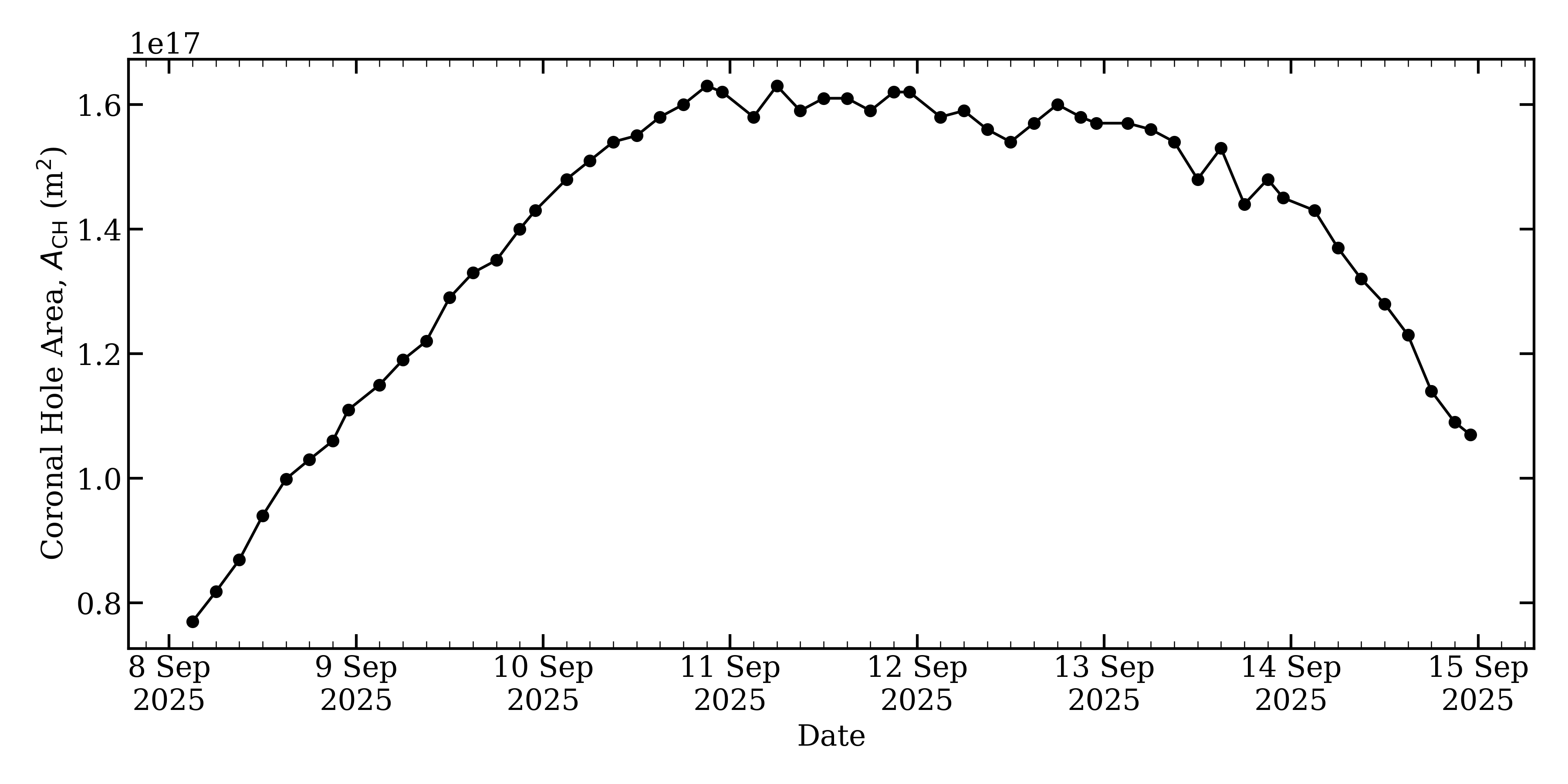}
	\caption{Temporal variation of the total coronal hole area from 8--14 September 2025. Values are expressed in m$^2$ and represent the average of 3-hourly measurements (8 datas per day) extracted using a 100 DN intensity mask.}
	\label{fig:area_result}
\end{figure*}

\subsection{Relationship Between $\langle I_{193\,\mathrm{\AA}}\rangle$,
$\Phi_{\text{unsigned}}$, and $A_{\text{CH}}$}

The observations from 8 to 14 September 2025 show a clear relationship between AIA EUV intensity ($\langle I_{193\,\mathrm{\AA}}\rangle$; DN), total unsigned magnetic flux ($\Phi_{\text{unsigned}}$; Mx), and coronal hole area ($A_{\mathrm{CH}}$; m$^2$). On 8 September, when a coronal hole was located near the eastern limb, the coronal hole area started at $7.70 \times 10^{16}$ m$^2$ at 03:00~UT and increased steadily to $1.11 \times 10^{17}$ m$^2$ by the end of the day at 23:00~UT. During the same period, AIA (DN) decreased from 61.23 DN to 54.27 DN, while HMI flux increased from $8.68 \times 10^{21}$ Mx to $1.25 \times 10^{22}$ Mx. This indicates that as the coronal hole grew, the EUV intensity decreased, while the magnetic flux increased.

From 9 to 10 September, the coronal hole area continued to grow, reaching a maximum of $1.63 \times 10^{17}$ m$^2$ on 10 September. AIA (DN) values continued to decline to about 45.14 DN, while HMI flux increased to approximately $1.86 \times 10^{22}$ Mx. Between 11 and 12 September, all three parameters remained relatively stable: the coronal hole area fluctuated around $1.54 \times 10^{17}$--$1.63 \times 10^{17}$ m$^2$, AIA (DN) stayed near 44--46 DN, and HMI flux remained around $1.79 \times 10^{22}$--$1.91 \times 10^{22}$ Mx. This suggests a temporary equilibrium in both EUV emission and magnetic flux during the mature phase of the coronal hole.

After 12 September, as the coronal hole moved toward the western limb, its area gradually decreased from $1.57 \times 10^{17}$ m$^2$ on 13 September to $1.07 \times 10^{17}$ m$^2$ on 14 September. During the same period, AIA (DN) increased from 46.62 DN to 52.61 DN, while HMI flux decreased from $1.83 \times 10^{22}$ Mx to $1.29 \times 10^{22}$ Mx. This opposite trend indicates that the reduction in coronal hole area is associated with higher EUV intensity and lower magnetic flux. These trends indicate a strong positive correlation between coronal hole area and unsigned magnetic flux (see \textbf{Equation \ref{eq:unsigned_flux}}), consistent with previous studies \cite{hofmeister2019photospheric, Lowder2015, Linker2021}.

\subsection{In Situ Signatures of the High-Speed Solar-Wind Stream}
\begin{figure*}[htp]
	\centering
	\includegraphics[width=0.9\textwidth]{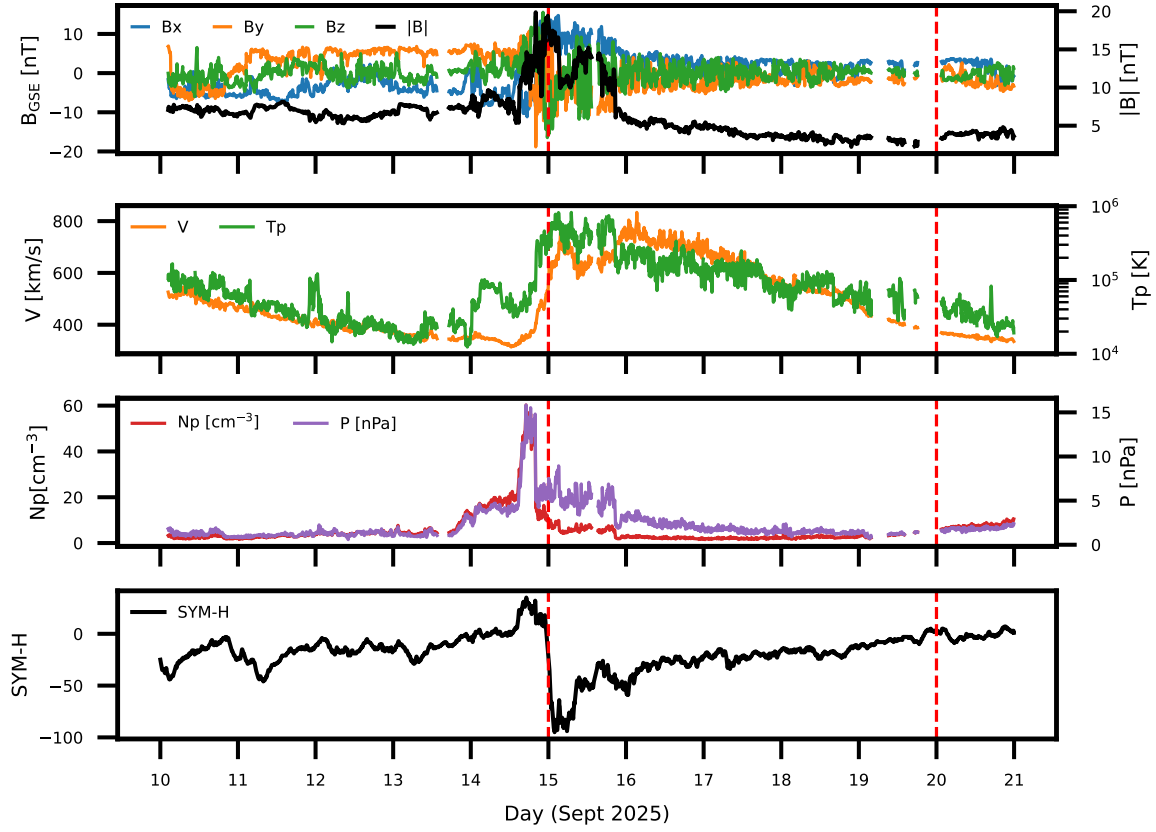}
	\caption{
		Panels (top to bottom) show $B_x$, $B_y$, $B_z$, and $|B|$; solar wind speed $V$ and proton temperature $T_p$; proton density $n_p$ and dynamic pressure $P$; and SYM-H. Red dashed lines mark the constant high-speed stream interval used for averaging. Data are from NASA GSFC OMNIWeb.
	}
	\label{fig:insuite_data}
\end{figure*}

The coronal hole observed from 8 to 14 September 2025 was associated with changes in the solar wind and interplanetary magnetic field measured at 1~AU between 10 and 20 September 2025. The SDO observations show a well-developed coronal hole with low AIA 193~\AA\ intensity, an unsigned magnetic flux ranging from $8.68\times10^{21}$ to $1.91\times10^{22}$~Mx, and an area ranging from $7.70\times10^{16}$ to $1.63\times10^{17}$~m$^{2}$. These properties are consistent with an open-field region capable of producing a high-speed solar-wind stream. Solar wind typically requires approximately 2--4 days to travel from the Sun to Earth \citep{Rotter2015}. A similar delay was observed between the coronal hole's central-meridian passage and the arrival of the high-speed stream at Earth.

On 8 and 9 September, when the coronal hole was near the eastern limb, the solar wind at 1~AU remained close to background conditions. The solar-wind speed was approximately $350$--$450~\mathrm{km\,s^{-1}}$, the proton density was about $5$--$10~\mathrm{cm^{-3}}$, and the magnetic-field strength was around $5$--$8~\mathrm{nT}$. During this period, the coronal hole was not yet well connected to the Sun--Earth direction. Therefore, the solar-wind stream associated with the coronal hole was expected to reach Earth only after the source region rotated closer to the central meridian \citep{rotter2012relation}.

Between 10 and 13 September, the coronal hole moved across the central part of the solar disk. Several days later, the in situ measurements showed a clear increase in solar-wind speed and proton temperature. The solar-wind speed reached approximately $600$--$750~\mathrm{km\,s^{-1}}$, while the proton temperature increased to about $(4$--$7)\times10^{5}$~K. The magnetic-field strength also increased to approximately $10$--$18~\mathrm{nT}$, with strong variations in the $B_x$, $B_y$, and $B_z$ components. Enhancements in proton density, dynamic pressure, and magnetic-field strength before the speed increase are consistent with a compressed interaction region at the leading edge of the high-speed stream. The decrease in the SYM-H index below $-50~\mathrm{nT}$ indicates a noticeable geomagnetic response \citep{tsurutani2006magnetic,hajra2022cir}.

The coronal hole reached its largest measured area and unsigned magnetic flux around 11--12 September. Its maximum unsigned magnetic flux was approximately $1.91\times10^{22}$~Mx. About four days later, the solar-wind speed reached nearly $750~\mathrm{km\,s^{-1}}$ at 1~AU. This timing supports an association between the central-meridian passage of the coronal hole and the observed high-speed solar-wind stream.

By 14 September, the coronal hole had moved toward the western limb. After 20 September, the solar-wind speed decreased to approximately $400$--$500~\mathrm{km\,s^{-1}}$, and the magnetic-field strength returned toward background levels. This behavior indicates the end of the high-speed-stream interval and a return to slower solar-wind conditions.

Overall, the observations show a consistent temporal relationship between the evolution of the butterfly-shaped coronal hole and the subsequent arrival of a high-speed solar-wind stream at Earth. These results support the established connection between coronal-hole magnetic structure and high-speed solar-wind flow \citep{Abramenko2009}.

\section{Conclusion}\label{sec:conclusion}
In this study, we examined the AIA 193~\AA\ mean intensity ($\langle I_{193,\mathrm{\AA}}\rangle$; DN), total unsigned magnetic flux ($\Phi_{\text{unsigned}}$; Mx), and coronal-hole area ($A_{\mathrm{CH}}$; $\mathrm{m}^2$) of a butterfly-shaped coronal hole observed from 8 to 14 September 2025 as it rotated across the solar disk from the eastern limb toward the western limb. By combining these remote-sensing observations with in situ solar-wind measurements at 1~AU, we draw the following conclusions:
\begin{enumerate}
   \item The coronal hole remained most stable near the disk center, where its area, total unsigned magnetic flux, and mean AIA 193~\AA\ intensity were nearly constant.

   \item The coronal-hole area and total unsigned magnetic flux varied together, indicating a strong relationship between the coronal-hole morphology and its underlying magnetic field.

   \item The mean AIA 193~\AA\ intensity varied inversely with the coronal-hole area and total unsigned magnetic flux.

   \item A high-speed solar-wind stream and its leading compressed interaction region were subsequently detected at 1~AU, producing a substantial geomagnetic response.
\end{enumerate}

\section*{Acknowledgments}
We acknowledge the SDO/AIA and SDO/HMI teams for providing the solar observations and the NASA GSFC/SPDF OMNIWeb service for the in-situ solar-wind data. The analysis code and supplementary materials are available at \url{https://github.com/alebhairab/Coronal-Hole-Detection}.

\bibliographystyle{unsrt}
\bibliography{references}

\end{multicols}

\end{document}